\documentclass[12pt]{article}
\begin{document}

\begin{center}
{\large \bf An Approximate Expression for the Large Angle Period of a Simple Pendulum}
\end{center}

\vspace{0.5in}

\begin{center}
{Rajesh R. Parwani\footnote{Email: parwani@nus.edu.sg}}

\vspace{0.5in}

{Department of Physics and \\
University Scholars Programme,\\}
{\#03-07, Old Administration Block,\\}
{National University of Singapore,\\}
{Kent Ridge, Singapore.}

\vspace{0.25in}


\end{center}

\begin{abstract}
A heuristic but pedagogical derivation is given of an explicit formula which accurately reproduces the period of a simple pendulum even for large amplitudes. The formula is compared with others in the literature.
\end{abstract}

\section{Introduction}

The dynamics of a simple pendulum for small amplitudes  is probably the most widely used example by educators to illustrate simple harmonic motion: Both the experiment and theory are accesible to college students. However the discussion of the large amplitude oscillations is rarely carried out, probably because there is no ``simple" analytical formula for the period. Over the years several approximation schemes have been developed   
to discuss the large angle oscillations and in this paper we provide yet another. It turns out that the formula derived here, which is quite accurate even for very large amplitudes, was obtained a long time ago by Ganley \cite{ganley} using a very different method. However, while the final result is the same, the derivation here is probably simpler and also directly linked to the usual derivation for the linearised pendulum. Indeed, the interested reader could potentially use the presentation here as the starting point for a systematic improvement  of the formula. Furthermore, an important difference between the result of Ganley and that obtained in this paper is mentioned at the end of the next section.

\section{Approximation Scheme}
The equation of motion for the simple pendulum can be derived by applying Newton's Second Law to motion along   
the tangential direction of the arc, giving

\begin{equation}
\ddot{\theta} + { g \over l} \sin{\theta} =0 \, ,
\end{equation}
where $l$ is the length of the pendulum, $g$  the acceleration due to gravity and $\theta$ the angular displacemnent. The usual linearised approximation involves replacing $\sin{\theta}$ by $\theta$. Let $ g/l = \omega_0^2$, so that the period, $T_0$, of a linear pendulum is $T_0= 2 \pi (\omega_{0}^{2})^{-1/2} = { 2 \pi \over \omega_0}$, which up to a constant of proportionality is the unique expression obtained by dimensional analysis. Indeed, by dimensional analysis the period of the non-linear pendulum must be of the form 

\begin{equation}
T_a = T_0 \ f(\alpha) \, ,
\end{equation}
where $\alpha$ is the amplitude of oscillations and $f(\alpha)$ a function to be determined. An approximate expression for $f(\alpha)$ can be obtained by first rewriting the equation of motion in the suggestive form

\begin{equation}
\ddot{\theta} + \omega_0^2 \left({\sin({\theta}) \over \theta}\right) \theta =0 \, . \label{rewrite}
\end{equation}

If one could treat the term in brackets as a constant, as it approximately is for small angles, then an attempted estimate of the period of the non-linear pendulum would be $ T_a \approx 2 \pi  \left( \omega_{0}^{2} {\sin(\theta) \over \theta} \right)^{-1/2}$. Of course this formula does not make much sense as it stands since the period cannot depend on the angular displacement: The varying $\theta$ should be replaced by some "average" value $\beta$ that depends on the amplitude $\alpha$. Thus we obtain  

\begin{equation}
f(\alpha) \approx   \left( \sin(\beta) \over \beta \right)^{-1/2}. \label{try1} 
\end{equation} 

Perturbative solutions \cite{pert} of the non-linear pendulum equation of motion give $f(\alpha) \approx 1 + { \alpha^2 \over 16}$. Comparing this with the perturbative expansion of the right-hand-side of Eq.(\ref{try1}) implies

\begin{equation}
{\alpha^2 \over 16} \approx {\beta^2 \over 12} \, ,
\end{equation}
or $\beta \approx {\sqrt{3} \alpha \over 2}$. Hence an first approximation for the period of a non-linear pendulum is 

\begin{equation}
T_{a1} \approx   T_0 \ \left( \sin({\sqrt{3} \alpha \over 2}) \over ({\sqrt{3} \alpha \over 2}) \right)^{-1/2}. \label{res1}
\end{equation} 
The quality of this approximation can be determined by comparing it to the exact value \cite{pert} given by

\begin{equation}
T_{ex} = {2 T_o \over \pi} \ K\left(\sin^2({\alpha \over 2})\right) \, , \label{ex}
\end{equation}
where 
\begin{equation}
K(m) = \int_{0}^{\pi/2} {dy \over (1-m \sin^2 y)^{1/2} } \, , 
\end{equation}
is the complete elliptic integral of the first kind.
The two expressions $T_{a1}/T_{0}$ and $T_{ex}/T_0$ are plotted in Figure(1) which shows that the approximation $T_{a1}$ gives an accuracy of $1\%$ for amplitudes as large as $\alpha \approx 2.25$ radians. It should be noted that the approximation always lies below the exact answer, but unlike the exact answer it does not diverge at $\alpha =\pi$.    

While the same formula for the {\it ratio} $T_{a1}/T_{0}$ was obtained by Ganley in an interesting paper \cite{ganley} using different methods, the approach presented here gives $T_{a1}$ explicitly and consequently in the limit of small angles the known expression $T_0$ is also recovered.

\section{Comparison with Other Approximations}

Perturbative solutions for  $f(\alpha)$ are known (or can be obtained by expansing the right-hand-side of (\ref{ex})). From the symmetry of the problem, one sees that only even powers of $\alpha$ can appear. A natural question is: ``How many terms of the series expansion must be kept to get a $1\%$ accuracy in the period for an amplitude $\alpha \approx 2.25$?" (that is, to compare with the approximate formula of the last section). A simple numerical exercise shows that all terms up to and including the eighth-order term must be taken into account.

In a recent paper Kidd and Fogg \cite{kidd} motivated the approximation $f(\alpha) \approx 1/\sqrt{\cos(\alpha/2)}$ which provides  a $1\%$ accuracy in the period for $\alpha$ up to $\pi /2$. Though the expression of Kidd and Fogg is accurate over a smaller range of amplitudes than the expression $T_{a1}$, it shares one interesting feature with the exact expression: A divergence at $\alpha = \pi$. This suggests that one generalise the method of the last section by trying an {\it ansatz} for $f(\alpha)$ that also diverges at the extreme amplitude  $\alpha = \pi$. One simple possibility, suggested by the form of Eq.(\ref{rewrite}) is 

\begin{equation}
f(\alpha) \approx \left( {\sin(\alpha) \over \alpha} \right)^{-\gamma} \,
\end{equation}   
where $\gamma$ is a positive constant to be determined.  By comparing the first nontrivial terms of the power-series of both sides of the last equation, one obtains $\gamma = {-3/8}$. The resulting expression 
\begin{equation}
T_{a2}= T_0 \ \left( {\sin(\alpha) \over \alpha} \right)^{-3/8} \, , 
\end{equation}
always exceeds the exact answer and has an accuracy of $1\%$ for amplitudes  up to $\alpha \approx 1.9$ radians.  Interestingly, the  
result $T_{a2}$ is precisely the expression of Molina \cite{molina} but obtained there with a slightly different motivation.

\section{Conclusion}

The approach presented in Sect.(2)  may possibly be viewed as a kind of ``adiabatic approximation" and perhaps improved on to give successively better approximations. The main result (\ref{res1}) is relatively simple and accurate to very large amplitudes.  While the {\it ratio} ${T_{a1} \over T_0}$ was previously obtained in \cite{ganley} using different arguments, here an explicit expression for $T_{a1}$  has been obtained.

\newpage

\input{epsf.sty}
{\bf Figure Caption}

Figure 1: Plot of $T_{ex}/T_0$ and $T_{a1}/T_0$ against the angular amplitude $\alpha$. The exact expression of course diverges for $\alpha = \pi$.\\

\vspace{2.0in}

\epsfbox{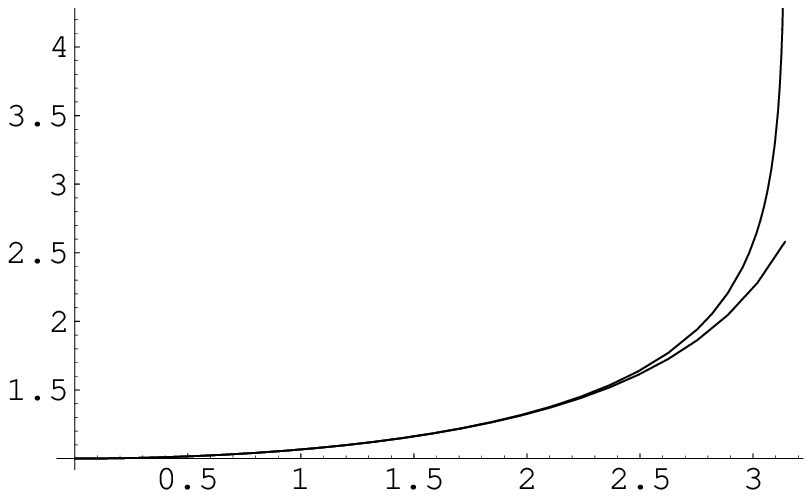}
\vspace{0.5in}

\end{document}